\documentclass[a4paper,11pt]{article}

\usepackage{jheppub,braket} 

\allowdisplaybreaks[3]

\begin{document}

  \title{Restoration of chiral symmetry in cold and dense Nambu--Jona-Lasinio model with tensor renormalization group}

  \author[a]{Shinichiro Akiyama,}
  \affiliation[a]{Graduate School of Pure and Applied Sciences, University of Tsukuba, Tsukuba, Ibaraki
    305-8571, Japan}
   \emailAdd{akiyama@het.ph.tsukuba.ac.jp}

  \author[b]{Yoshinobu Kuramashi,}
  \affiliation[b]{Center for Computational Sciences, University of Tsukuba, Tsukuba, Ibaraki
    305-8577, Japan}
  \emailAdd{kuramasi@het.ph.tsukuba.ac.jp}

  \author[c]{Takumi Yamashita,}
  \affiliation[c]{Faculty of Engineering, Information and Systems, University of Tsukuba, Tsukuba, Ibaraki
    305-8573, Japan}  
  \emailAdd{yamasita@ccs.tsukuba.ac.jp}

  \author[b]{Yusuke Yoshimura}
  \emailAdd{yoshimur@ccs.tsukuba.ac.jp}

\abstract{

We analyze the chiral phase transition of the Nambu--Jona-Lasinio model in the cold and dense region on the lattice, developing the Grassmann version of the anisotropic tensor renormalization group algorithm. The model is formulated with the Kogut--Susskind fermion action. We use the chiral condensate as an order parameter to investigate the restoration of the chiral symmetry. The first-order chiral phase transition is clearly observed in the dense region at vanishing temperature with $\mu/T\sim O(10^3)$ on a large volume of $V=1024^4$. We also present the results for the equation of state.
}
\date{\today}

\preprint{UTHEP-752, UTCCS-P-134}

\maketitle

\section{Introduction}
\label{sec:intro}

The phase structure and the equation of state for QCD at finite temperature and density are essential ingredients to understand the evolution and the current state of the universe quantitatively. Although the lattice QCD simulation has been expected to be an ideal tool to investigate the non-perturbative aspects of QCD, it has not been successful to reveal the nature of QCD at finite density. This is due to the sign problem caused by the introduction of the chemical potential in the lattice QCD simulations based on the Monte Carlo algorithm, see, {\it e.g.}, Ref.~\cite{deForcrand:2010ys}.
  
The tensor renormalization group (TRG) method, which was originally proposed by Levin and Nave to study two-dimensional ($2d$) classical spin models in 2007 \cite{Levin:2006jai}, has several superior features over the Monte Carlo method.\footnote{In this paper the TRG method or the TRG approach refers to not only the original numerical algorithm proposed by Levin and Nave but also its extensions \cite{PhysRevB.86.045139,Adachi:2019paf,Kadoh:2019kqk,Shimizu:2014uva,Sakai:2017jwp}.} Firstly, the TRG method, and also other tensor network methods, are free from the sign problem. This virtue was confirmed by successful application of these methods to various $2d$ quantum field theories which contain the sign problem \cite{Shimizu:2014uva,Shimizu:2014fsa,Shimizu:2017onf,Takeda:2014vwa,Kadoh:2018hqq,Kadoh:2019ube,Kuramashi:2019cgs}.\footnote{The TRG study of the sign problem was also performed by introducing the complex coupling or the chemical potential to the $2d$ classical $O(2)$ spin model~\cite{Denbleyker:2013bea,Yang:2015rra}.}  Moreover, the authors have successfully employed the anisotropic TRG (ATRG) algorithm to analyze the Bose condensation in the $4d$ complex $\phi^4$ theory at finite density \cite{Akiyama:2020ntf}. Secondly, the computational cost depends on the system size only logarithmically. This is an essential ingredient to enable us to take the thermodynamic limit at vanishing temperature. Thirdly, the Grassmann version of the TRG method, which was developed by some of the authors \cite{Shimizu:2014uva,Sakai:2017jwp,Yoshimura:2017jpk}, allows direct manipulation of the Grassmann variables. Fourthly, we can obtain the partition function or the path-integral itself. In the thermodynamic limit, the pressure is directly related to the thermodynamic potential so that the equation of state can be easily obtained with the TRG method.

\begin{figure}[htbp]
	\centering
	\includegraphics[width=0.8\hsize,bb=0 0 720 540]{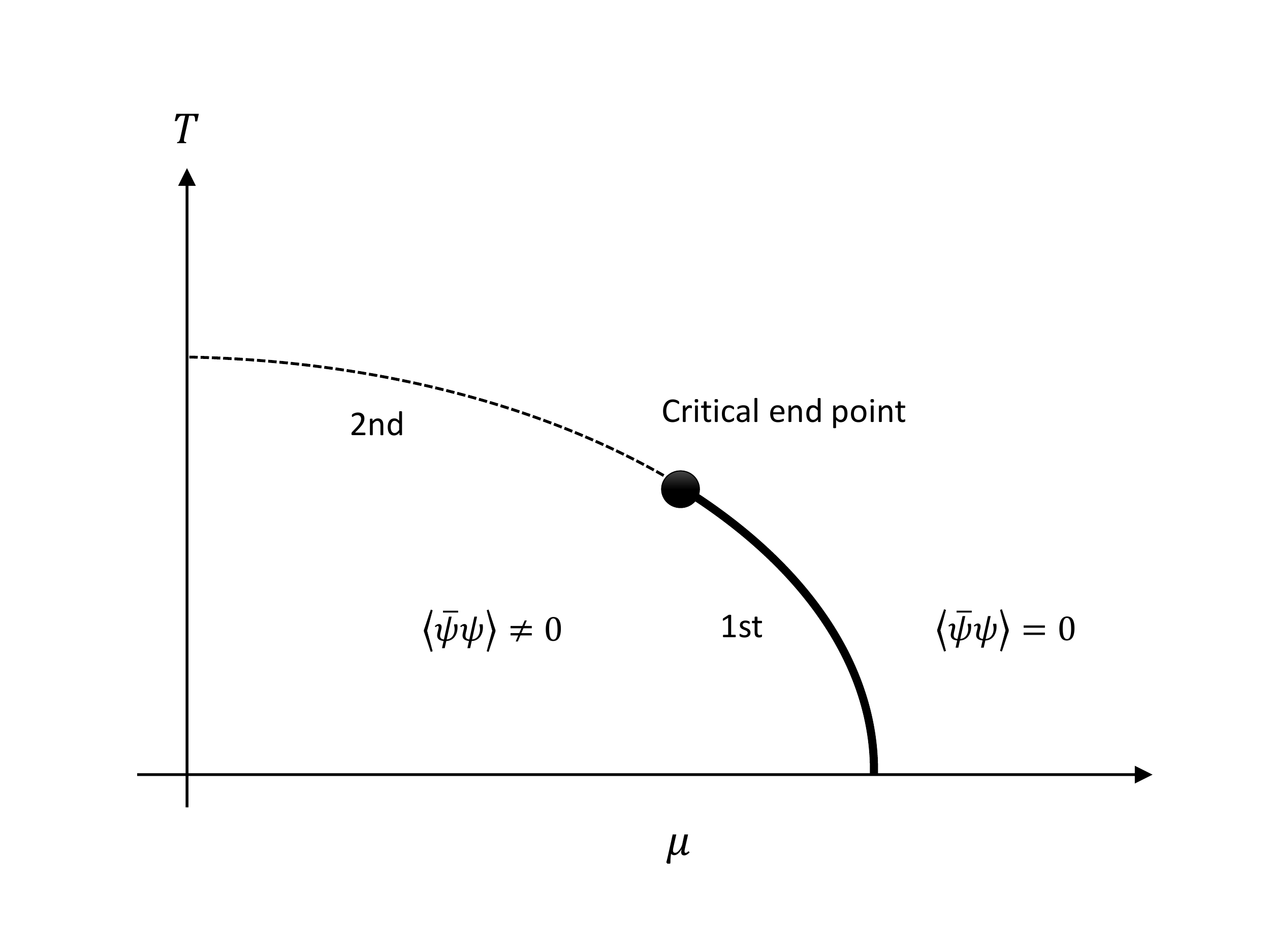}
  	\caption{Schematic view of expected phase diagram of the NJL model on the $T$-$\mu$ plane. Solid and broken curves represent the first- and second-order phase transitions, respectively. Closed circle denotes the critical end point (CEP) where the first-order phase transition line terminates.}
  	\label{fig:pd_anal}
\end{figure}

In this paper we investigate the phase structure of the Nambu--Jona-Lasinio (NJL) model \cite{Nambu:1961tp,Nambu:1961fr} at finite temperature $T$ and chemical potential $\mu$ on the lattice, developing the Grassmann version of the ATRG algorithm. The Lagrangian of the NJL model in the continuum is defined as follows:
\begin{align}
	\label{eq:njl_cont}
	{\cal L}= {\bar \psi}(x)\gamma_\nu\partial_\nu \psi(x) 
 	-g_0\left\{({\bar \psi}(x)\psi(x))^2+({\bar \psi}(x){\rm i}\gamma_5\psi(x))^2 \right\},
\end{align}
which has the U(1) chiral symmetry with $\psi(x) \rightarrow {\rm e}^{{\rm i}\alpha\gamma_5}\psi(x)$ and ${\bar \psi}(x) \rightarrow {\bar \psi}(x){\rm e}^{{\rm i}\alpha\gamma_5}$. This is an effective theory of QCD which describes the dynamical chiral symmetry breaking: once the strength of the coupling constant $g_0$ exceeds a certain critical value the system generates a non-trivial vacuum with $\braket{ {\bar \psi}(x)\psi(x)} \ne 0$. The chiral phase structure of the NJL model on the $T$-$\mu$ plane is discussed by some analytical methods, $e.g.$, the mean-field approximation (MFA) \cite{Buballa:2003qv} and the functional renormalization group (FRG) \cite{Aoki:2017rjl}. Figure~\ref{fig:pd_anal} shows a schematic view of the expected phase structure, whose characteristic feature is the first-order chiral phase transition in the dense region at very low temperature \cite{Asakawa:1989bq}. This phase transition is our primary target to investigate, employing the chiral condensate $\braket{ {\bar \psi}(x)\psi(x)}$ as an order parameter. Since the chiral symmetry plays a crucial role in this study, we use the Kogut--Susskind fermion to formulate the NJL model on the lattice.
 The analysis of the phase structure with the TRG method would help us understand the thermodynamic properties of dense QCD. 

This paper is organized as follows. In Sec.~\ref{sec:method} we explain the formulation of the lattice NJL model with the Kogut--Susskind fermion and the algorithmic details of the Grassmann ATRG (GATRG). In Sec.~\ref{sec:results}, we compare the numerical results with the analytical ones in the heavy dense limit as a benchmark before numerical results for the chiral condensate and the equation of state are presented. Section~\ref{sec:summary} is devoted to summary and outlook.

\section{Formulation and numerical  algorithm}
\label{sec:method}

\subsection{NJL model on the lattice}
\label{subsec:njl}

We use the Kogut--Susskind fermion to formulate the NJL model on the lattice. Following Refs.~\cite{Lee:1987eg,Booth:1989ms}, we define the model at finite chemical potential $\mu$ as
\begin{align}
	\label{eq:njl_lat}
	S=&\frac{1}{2}a^3\sum_{n\in\Lambda}\sum_{\nu=1}^{4} \eta_\nu(n)\left[{\rm e}^{\mu a\delta_{\nu,4}}{\bar \chi}(n)\chi(n+{\hat\nu})-{\rm e}^{-\mu a\delta_{\nu,4}}{\bar \chi}(n+{\hat\nu})\chi(n)\right]\nonumber\\
	&+ma^4\sum_{n\in\Lambda}{\bar \chi}(n)\chi(n)-g_{0}a^4\sum_{n\in\Lambda}\sum_{\nu=1}^{4}{\bar \chi}(n)\chi(n){\bar \chi}(n+{\hat\nu})\chi(n+{\hat\nu}),
\end{align}
where $n=(n_1,n_2,n_3,n_4)(\in\mathbb{Z}^4)$ specifies a position in the lattice $\Lambda$, with the lattice spacing $a$. $\chi(n)$ and $\bar{\chi}(n)$ are Grassmann-valued fields without the Dirac structure. Since they describe the Kogut--Susskind fermions, $\chi(n)$ and $\bar{\chi}(n)$ are single-component Grassmann variables. $\eta_{\nu}(n)$ is the staggered sign function defined by $\eta_\nu(n)=(-1)^{n_1+\cdots +n_{\nu-1}}$ with $\eta_1(n)=1$. The partition function is defined in the ordinal manner:
\begin{align}
\label{eq:partition}
	Z=\int\left(\prod_{n\in\Lambda}{\rm d}\chi(n){\rm d}\bar{\chi}(n)\right){\rm e}^{-S}.
\end{align}
For vanishing mass $m$, Eq.~\eqref{eq:njl_lat} is invariant under the following continuous chiral transformation:
\begin{align}
	\chi(n)&\rightarrow {\rm e}^{{\rm i}\alpha\epsilon(n)}\chi(n),\\
	{\bar \chi}(n)&\rightarrow {\bar \chi}(n){\rm e}^{{\rm i}\alpha\epsilon(n)}
\end{align}
with $\alpha\in\mathbb{R}$ and $\epsilon(n)=(-1)^{n_1+n_2+n_3+n_4}$. 

\subsection{Tensor network representation}
\label{subsec:tnrep}

\begin{figure}[htbp]
	\centering
	\includegraphics[width=1.0\hsize,bb= 0 0 960 549]{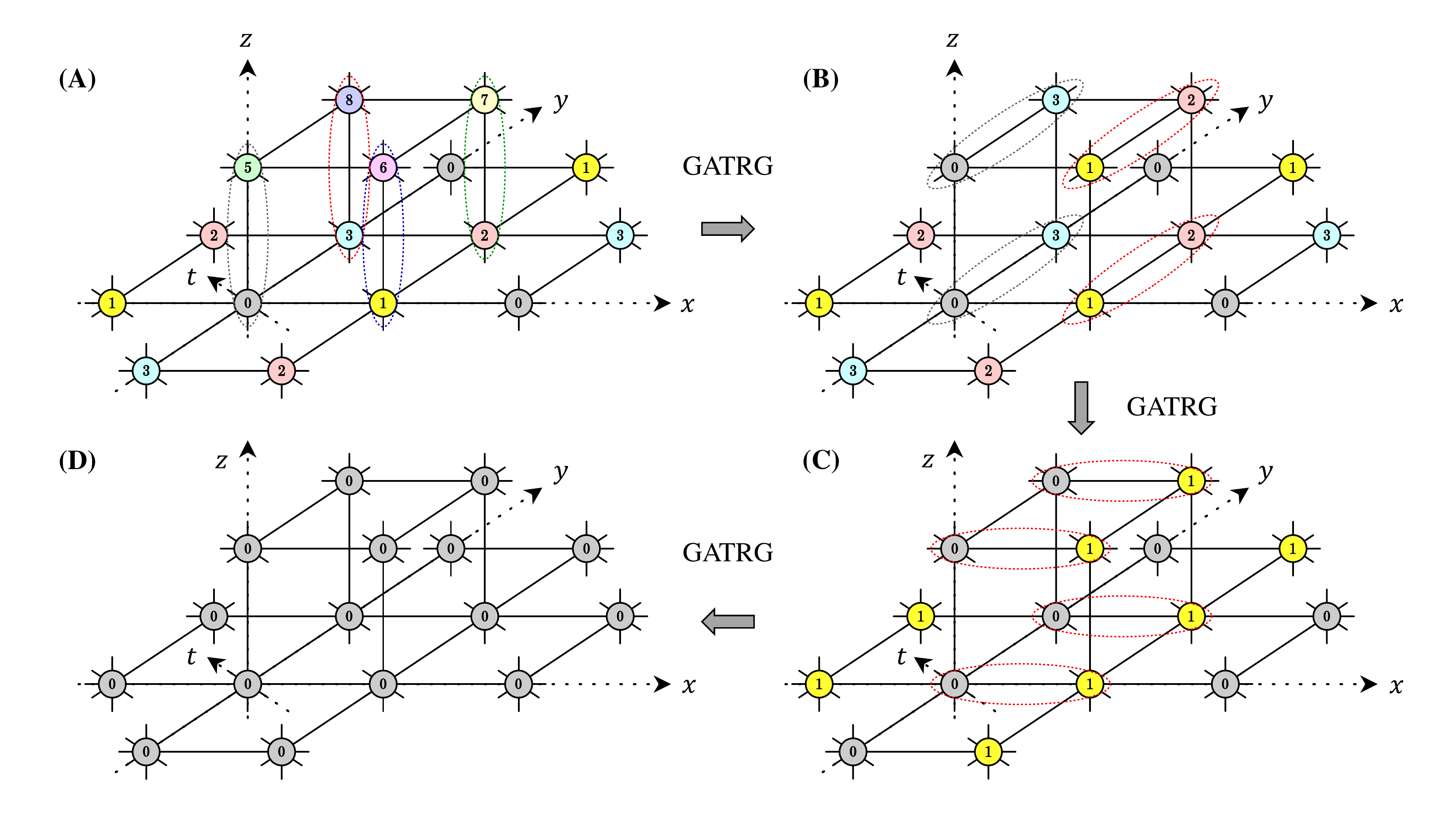}
  	\caption{Schematic illustration of the coarse-graining with the GATRG. Coordinate axes in $\Lambda$ are shown as dotted lines with arrows. Different numbers are assigned to specify different tensors. (A) Initial tensor network in Eq.~\eqref{eq:network}. Eight types of tensor are located at a spatial unit cube. A periodic structure is explicitly shown on $x$-$y$ plane. The tensor network is uniform in $t$-direction. (B) The first coarse-graining along $z$-direction reduces types of tensor from eight to four. The tensor network becomes uniform in $t$-, $z$-directions. (C) The second coarse-graining along $y$-direction makes the structure uniform except in $x$-direction. (D) Totally uniform tensor network is obtained by the third coarse-graining along $x$-direction. In the following coarse-graining steps, the structure is invariant.}
  	\label{fig:network}
\end{figure}

We introduce the tensor network representation for Eq.~\eqref{eq:partition} in a similar way with Refs.~\cite{Takeda:2014vwa,Kadoh:2018hqq}.\footnote{See Ref.~\cite{PhysRevD.101.094509} for a different TRG approach with the Kogut-Susskind fermion, where the TRG procedure is applied to the Schwinger model after integrating out the fermion fields analytically.} Hereafter, we set $a=1$ for simplicity. Firstly, we expand the local Boltzmann weights in the following manners to decompose the nearest-neighbor interactions:
\begin{align}
\label{eq:forward}
	&{\rm exp}\left[-\frac{{\rm e}^{\mu\delta_{\nu,4}}}{2}\eta_{\nu}(n)\bar{\chi}(n)\chi(n+\hat{\nu})\right]=\sum_{i_{\nu,1}(n)=0}^{1}\int\nonumber\\
	&\left(\frac{{\rm e}^{\frac{\mu}{2}\delta_{\nu,4}}}{\sqrt{2}}\eta_{\nu}(n)\bar{\chi}(n){\rm d}\Phi_{\nu}(n)\cdot\frac{{\rm e}^{\frac{\mu}{2}\delta_{\nu,4}}}{\sqrt{2}}\chi(n+\hat{\nu}){\rm d}\bar{\Phi}_{\nu}(n+\hat{\nu})\cdot\bar{\Phi}_{\nu}(n+\hat{\nu})\Phi_{\nu}(n)\right)^{i_{\nu,1}(n)},
\end{align}
\begin{align}
\label{eq:backward}
	&{\rm exp}\left[\frac{{\rm e}^{-\mu\delta_{\nu,4}}}{2}\eta_{\nu}(n)\bar{\chi}(n+\hat{\nu})\chi(n)\right]=\sum_{i_{\nu,2}(n)=0}^{1}\int\nonumber\\
	&\left(\frac{{\rm e}^{-\frac{\mu}{2}\delta_{\nu,4}}}{\sqrt{2}}\eta_{\nu}(n)\chi(n){\rm d}\Psi_{\nu}(n)\cdot\frac{{\rm e}^{-\frac{\mu}{2}\delta_{\nu,4}}}{\sqrt{2}}\bar{\chi}(n+\hat{\nu}){\rm d}\bar{\Psi}_{\nu}(n+\hat{\nu})\cdot\bar{\Psi}_{\nu}(n+\hat{\nu})\Psi_{\nu}(n)\right)^{i_{\nu,2}(n)},
\end{align}
\begin{align}
\label{eq:four-fermi}
	&{\rm e}^{g_{0}{\bar \chi}(n)\chi(n){\bar \chi}(n+{\hat\nu})\chi(n+{\hat\nu})}=
	&\sum_{i_{\nu,3}(n)=0}^{1}\left(\sqrt{g_{0}}{\bar \chi}(n)\chi(n)\cdot\sqrt{g_{0}}{\bar \chi}(n+{\hat\nu})\chi(n+{\hat\nu})\right)^{i_{\nu,3}(n)}.
\end{align}
Secondly, integrating out $\chi$ and $\bar{\chi}$ at each lattice site $n$, we define
\begin{align}
	&\mathcal{T}_{n;i_{4}(n)i_{1}(n)i_{2}(n)i_{3}(n)i_{4}(n-\hat{4})i_{1}(n-\hat{1})i_{2}(n-\hat{2})i_{3}(n-\hat{3})}\nonumber\\
	&=\int{\rm d}\chi{\rm d}\bar{\chi}~{\rm e}^{-m\bar{\chi}\chi}\prod_{\nu=1}^{4}\left(\frac{{\rm e}^{\frac{\mu}{2}\delta_{\nu,4}}}{\sqrt{2}}\eta_{\nu}(n)\bar{\chi}{\rm d}\Phi_{\nu}(n)\right)^{i_{\nu,1}(n)}\left(\frac{{\rm e}^{\frac{\mu}{2}\delta_{\nu,4}}}{\sqrt{2}}\chi{\rm d}\bar{\Phi}_{\nu}(n)\right)^{i_{\nu,1}(n-\hat{\nu})}\nonumber\\
	&\left(\frac{{\rm e}^{-\frac{\mu}{2}\delta_{\nu,4}}}{\sqrt{2}}\eta_{\nu}(n)\chi{\rm d}\Psi_{\nu}(n)\right)^{i_{\nu,2}(n)}\left(\frac{{\rm e}^{-\frac{\mu}{2}\delta_{\nu,4}}}{\sqrt{2}}\bar{\chi}{\rm d}\bar{\Psi}_{\nu}(n)\right)^{i_{\nu,2}(n-\hat{\nu})}\left(\sqrt{g_{0}}{\bar \chi}\chi\right)^{i_{\nu,3}(n)}\left(\sqrt{g_{0}}{\bar \chi}\chi\right)^{i_{\nu,3}(n-\hat{\nu})}\nonumber\\
	&\left(\bar{\Phi}_{\nu}(n+\hat{\nu})\Phi_{\nu}(n)\right)^{i_{\nu,1}(n)}\left(\bar{\Psi}_{\nu}(n+\hat{\nu})\Psi_{\nu}(n)\right)^{i_{\nu,2}(n)}.
\end{align}
This serves as a change of variables from $\chi,\bar{\chi}$ to the integer-valued fields $i_{\nu}=(i_{\nu,p})_{p=1,2,3}$ and alternative Grassmann variables $\Phi_{\nu},\Psi_{\nu}$. Renaming $x=i_{1},y=i_{2},z=i_{3},t=i_{4}$, Eq.~\eqref{eq:partition} is expressed in the form,
\begin{align}
\label{eq:network}
	Z=\sum_{\{t,x,y,z\}}\int\prod_{n\in\Lambda}\mathcal{T}_{n;txyzt'x'y'z'},
\end{align}
which is the tensor network representation of this model.\footnote{In Eq.~\eqref{eq:network}, we omit arguments in tensor indices and introduce shorthand notations such as $x'=x(n-\hat{1}),y'=y(n-\hat{2}),z'=z(n-\hat{3}),t'=t(n-\hat{4})$} In current construction, $\mathcal{T}_{n;txyzt'x'y'z'}$ is factorized as
\begin{align}
\label{eq:structure}
	\mathcal{T}_{n;txyzt'x'y'z'}=I_{n;txyzt'x'y'z'}S_{n;txyzt'x'y'z'}\mathcal{G}_{n;txyzt'x'y'z'}.
\end{align}
$I_{n;txyzt'x'y'z'}$ denotes the contributions from the integration over $\chi(n)$ and $\bar{\chi}(n)$. A straightforward calculation shows
\begin{align}
\label{eq:i-tensor}
	&I_{n;txyzt'x'y'z'}=\nonumber\\
	&(-1)^{n_1(y_1+y_2+z_1+z_2+t_1+t_2)+n_2(z_1+z_2+t_1+t_2)+n_3(t_1+t_2)}\nonumber\\
	&\left(\frac{1}{\sqrt{2}}\right)^{t_1+t_2+x_1+x_2+y_1+y_2+z_1+z_2+t'_1+t'_2+x'_1+x'_2+y'_1+y'_2+z'_1+z'_2}\sqrt{g_0}^{t_3+x_3+y_3+z_3+t'_3+x'_3+y'_3+z'_3}\nonumber\\
	&{\rm e}^{\frac{\mu}{2}(t_1-t_2+t'_1-t'_2)}\left[-m\bar{\Delta}_{txyzt'x'y'z',0}\Delta_{txyzt'x'y'z',0}+\bar{\Delta}_{txyzt'x'y'z',1}\Delta_{txyzt'x'y'z',1}\right],
\end{align}
where
\begin{align}
	\bar{\Delta}_{txyzt'x'y'z',q}=\delta_{t_1+t_3+x_1+x_3+y_1+y_3+z_1+z_3+t'_2+t'_3+x'_2+x'_3+y'_2+y'_3+z'_2+z'_3,q},\\
	\Delta_{txyzt'x'y'z',q}=\delta_{t_2+t_3+x_2+x_3+y_2+y_3+z_2+z_3+t'_1+t'_3+x'_1+x'_3+y'_1+y'_3+z'_1+z'_3,q},
\end{align}
with $q=0,1$. $\bar{\Delta},\Delta$ are derived from $\bar{\chi}$-, $\chi$-integration, respectively. The second line in Eq.~\eqref{eq:i-tensor} comes from the staggered sign factor $\eta_{\nu}(n)$. Consequently, $I_{n;txyzt'x'y'z'}$ does depend on $n\in\Lambda$. Eq.~\eqref{eq:i-tensor} tells us that this tensor network is uniform in $t$-direction, but has some periodic structure in $x$-,$y$-,$z$-directions. This periodicity corresponds to the parity of the spatial lattice site ${\bf n}=(n_1,n_2,n_3)$. A graphical representation of Eq.~\eqref{eq:network} is shown in Fig.~\ref{fig:network}~(A). As a result of Eqs.~\eqref{eq:forward} and \eqref{eq:backward},  some Grassmann variables are allowed to exist in $\mathcal{T}_{n;txyzt'x'y'z'}$. These Grassmann variables have been denoted by $\mathcal{G}_{n;txyzt'x'y'z'}$ in Eq.~\eqref{eq:structure}. Some sign can arise reflecting on how we have arranged these Grassmann variables in $\mathcal{G}_{n;txyzt'x'y'z'}$ and we have set this sign $S_{n;txyzt'x'y'z'}$ in Eq.~\eqref{eq:structure}. We now assume that
\begin{align}
\label{eq:grassmann-order}
	&\mathcal{G}_{n;txyzt'x'y'z'}=\nonumber\\
	&{\rm d}\Phi_{4}^{t_1}{\rm d}\Psi_{4}^{t_2}{\rm d}\Phi_{1}^{x_1}{\rm d}\Psi_{1}^{x_2}{\rm d}\Phi_{2}^{y_1}{\rm d}\Psi_{2}^{y_2}{\rm d}\Phi_{3}^{z_1}{\rm d}\Psi_{3}^{z_2}{\rm d}\bar{\Psi}_{4}^{t'_2}{\rm d}\bar{\Phi}_{4}^{t'_1}{\rm d}\bar{\Psi}_{1}^{x'_2}{\rm d}\bar{\Phi}_{1}^{x'_1}{\rm d}\bar{\Psi}_{2}^{y'_2}{\rm d}\bar{\Phi}_{2}^{y'_1}{\rm d}\bar{\Psi}_{3}^{z'_2}{\rm d}\bar{\Phi}_{3}^{z'_1}\nonumber\\
	&\left(\bar{\Phi}_{4}(n+\hat{4})\Phi_{4}(n)\right)^{t_1}\left(\bar{\Psi}_{4}(n+\hat{4})\Psi_{4}(n)\right)^{t_2}\left(\bar{\Phi}_{1}(n+\hat{1})\Phi_{1}(n)\right)^{x_1}\left(\bar{\Psi}_{1}(n+\hat{1})\Psi_{1}(n)\right)^{x_2}\nonumber\\
	&\left(\bar{\Phi}_{2}(n+\hat{2})\Phi_{2}(n)\right)^{y_1}\left(\bar{\Psi}_{2}(n+\hat{2})\Psi_{2}(n)\right)^{y_2}\left(\bar{\Phi}_{3}(n+\hat{3})\Phi_{3}(n)\right)^{z_1}\left(\bar{\Psi}_{3}(n+\hat{3})\Psi_{3}(n)\right)^{z_2},
\end{align}
where all the Grassmann measures depend on $n$ and their arguments are omitted. According to this arrangement, $S_{n;txyzt'x'y'z'}$ is given by
\begin{align}
	S_{n;txyzt'x'y'z'}&=(-1)^{t_1(t_2+x_2+y_2+z_2)+x_1(x_2+y_2+z_2)+y_1(y_2+z_2)+z_1z_2}\nonumber\\
	&~\quad(-1)^{t'_2(t'_1+x'_1+y'_1+z'_1)+x'_2(x'_1+y'_1+z'_1)+y'_2(y'_1+z'_1)+z'_2z'_1}\nonumber\\
	&~\quad(-1)^{(t_1+t_2+x_1+x_2+y_1+y_2+z_1+z_2)(t'_1+x'_1+y'_1+z'_1)}.
\end{align}

\subsection{Grassmann ATRG}
\label{subsec:gatrg}

\subsubsection{Procedure of the algorithm}
\label{subsubsec:gatrg}

We now formulate Grassmann ATRG (GATRG) algorithm to coarse grain the tensor network defined by Eq.~\eqref{eq:network}. The basic idea is that we combine the ATRG procedure to compress $I_{n;txyzt'x'y'z'}S_{n;txyzt'x'y'z'}$ in Eq.~\eqref{eq:structure} with the Grassmann HOTRG (GHOTRG) procedure to deal with $\mathcal{G}_{n;txyzt'x'y'z'}$ in Eq.~\eqref{eq:structure}. 

Let us consider the coarse-graining along $z$-direction. Our aim is to construct a kind of block-spin transformation, $\mathcal{T}_{n+\hat{3}}\cdot\mathcal{T}_{n}\mapsto\mathcal{T}'$. We start from constructing the transformation $\mathcal{G}_{n+\hat{3}}\cdot\mathcal{G}_{n}\mapsto\mathcal{G}'$, employing the GHOTRG procedure demonstrated in Refs.~\cite{Sakai:2017jwp,Yoshimura:2017jpk}. A straightforward extension to the $4d$ system gives us
\begin{align}
\label{eq:new_g_tensor}
	\mathcal{G}'_{\tilde{t}\tilde{x}\tilde{y}z\tilde{t}'\tilde{x}'\tilde{y}'z'}&=\sigma_{n+\hat{3},n}{\rm d}\eta^{\tilde{t}}{\rm d}\xi^{\tilde{x}}{\rm d}\theta^{\tilde{y}}{\rm d}\Phi_{3}^{z_1}{\rm d}\Psi_{3}^{z_2}{\rm d}\bar{\eta}^{\tilde{t}'}{\rm d}\bar{\xi}^{\tilde{x}'}{\rm d}\bar{\theta}^{\tilde{y}'}{\rm d}\bar{\Psi}_{3}^{z'_2}{\rm d}\bar{\Phi}_{3}^{z'_1}\nonumber\\
	&\quad(\bar{\eta}\eta)^{\tilde{t}}(\bar{\xi}\xi)^{\tilde{x}}(\bar{\theta}\theta)^{\tilde{y}}\left(\bar{\Phi}_{3}(n+\hat{3})\Phi_{3}(n)\right)^{z_1}\left(\bar{\Psi}_{3}(n+\hat{3})\Psi_{3}(n)\right)^{z_2}.
\end{align}
One can confirm this expression by integrating out some of the Grassmann variables in $\mathcal{G}_{n+\hat{3}}\cdot\mathcal{G}_{n}$ with the help of some shifting technique.\footnote{See Ref.~\cite{Sakai:2017jwp} for more details.} From now on, we call the exponent of Grassmann number as the fermion index. In Eq.~\eqref{eq:new_g_tensor}, new fermion indices, $\tilde{t},\tilde{x},\tilde{y},\tilde{t}',\tilde{x}',\tilde{y}'$, are introduced with new Grassmann variables, $\eta,\xi,\theta,\bar{\eta},\bar{\xi},\bar{\theta}$. The resulting sign factor $\sigma_{n+\hat{3},n}$ in Eq.~\eqref{eq:new_g_tensor} is
\begin{align}
\label{eq:sign}
	\sigma_{n+\hat{3},n}&=(-1)^{z_1(n)+z_2(n)}\nonumber\\
	&~\quad(-1)^{[t_1(n+\hat{3})+t_2(n+\hat{3})][t_1(n)+t_2(n)]+[t_1(n+\hat{3})+t_2(n+\hat{3})+x_1(n+\hat{3})+x_2(n+\hat{3})][x_1(n)+x_2(n)]}\nonumber\\
	&~\quad(-1)^{[t_1(n+\hat{3})+t_2(n+\hat{3})+x_1(n+\hat{3})+x_2(n+\hat{3})+y_1(n+\hat{3})+y_2(n+\hat{3})][y_1(n)+y_2(n)]}\nonumber\\
	&~\quad(-1)^{[y'_1(n+\hat{3})+y'_2(n+\hat{3})][y'_1(n)+y'_2(n)]+[x'_1(n+\hat{3})+x'_2(n+\hat{3})+y'_1(n+\hat{3})+y'_2(n+\hat{3})][x'_1(n)+x'_2(n)]}\nonumber\\
	&~\quad(-1)^{[t'_1(n+\hat{3})+t'_2(n+\hat{3})+x'_1(n+\hat{3})+x'_2(n+\hat{3})+y'_1(n+\hat{3})+y'_2(n+\hat{3})][t'_1(n)+t'_2(n)]}.
\end{align}
We now move on to the ATRG procedure for $I_{n;txyzt'x'y'z'}S_{n;txyzt'x'y'z'}$. In the following, we set $T_{n;txyzt'x'y'z'}=I_{n;txyzt'x'y'z'}S_{n;txyzt'x'y'z'}$. Incorporating with $\sigma_{n+\hat{3},n}$ in Eq.~\eqref{eq:new_g_tensor}, a total block-spin transformation, $\mathcal{T}_{n+\hat{3}}\cdot\mathcal{T}_{n}\mapsto\mathcal{T}'$, is accomplished. In other words, what we have to formulate is a block-spin transformation, $\sigma_{n+\hat{3},n}\cdot T_{n+\hat{3}}\cdot T_{n}\mapsto T'$. The sign factor $\sigma_{n+\hat{3},n}$ consists of two types of index; ones are $z_1,z_2$ living on the coarse-graining direction and the others are $t_1,t_2,x_1,x_2,y_1,y_2$ living on the other directions. We deal with them in a separate way. Let $\sigma^{(\rm CG)}_n=(-1)^{z_1(n)+z_2(n)}$ and $\sigma^{(\rm NCG)}_{n+\hat{3},n}=\sigma_{n+\hat{3},n}/\sigma^{(\rm CG)}_n$. Then $\sigma^{(\rm CG)}_n$ is included in the swapping bond part of the ATRG.\footnote{See Refs.\cite{Adachi:2019paf,Oba:2019csk} for more details about the swapping bond part and the squeezers used in the ATRG algorithm.} This is quite natural because one has to contract the tensors with respect to the index $z$ in the swapping step of the ATRG. On the other hand, $\sigma^{(\rm NCG)}_{n+\hat{3},n}$ is handled both in finding squeezers and in contracting these squeezers and local tensors. Modifying these procedures in the ATRG, the GATRG is formulated.

The coarse-graining along $z$-direction is followed by the series of coarse-graining along $y$-, $x$-, and $t$-directions. Fig.~\ref{fig:network} shows a schematic picture of the first three times of coarse-graining. Though the original tensor network in Eq.~\eqref{eq:network} consists of several types of local tensor, the GATRG reduces them under a sequential coarse-graining process and we obtain a uniform tensor network in all directions.

As a final supplement, we comment on another implementation for GATRG. It is also possible to coarse-grain $\mathcal{G}_{n;txyzt'x'y'z'}$ following the philosophy of the ATRG. $\mathcal{G}_{n;txyzt'x'y'z'}$ can be decomposed by introducing some extra Grassmann variables in the swapping bond part, based on the same idea in the Grassmann TRG \cite{Shimizu:2014uva,Takeda:2014vwa}. This implementation must reproduce the same result with the GATRG explained above if no finite bond dimension is introduced. However, the results obtained with the finite bond calculation are possibly different because these two GATRGs assume non-identical cost functions in optimization. We have numerically confirmed that this another GATRG also works, applying it to evaluate the tensor network in Eq.~\eqref{eq:network}. We have also found that the deviation between resulting thermodynamic potentials obtained by two types of GATRG tends to be smaller as the bond dimension is increased. 

\subsubsection{Some techniques}
\label{subsubsec:technique}

Eq.~\eqref{eq:i-tensor} reveals that $\mathcal{T}$ takes a finite value if and only if its Grassmann parity is even. This feature can be understood as a kind of $\mathbb{Z}_2$ symmetry, which enables us to introduce the block diagonal representation for some tensors treated in the GATRG algorithm. We carry out the singular value decomposition (SVD) in the swapping bond part and the higher-order SVD in finding squeezers under the block diagonal representation for corresponding tensors. This blocking technique is of essential importance because it naturally defines the fermion indices introduced in Eq.~\eqref{eq:new_g_tensor}. For instance, in order to find the squeezers in coarse-graining along $\nu$-direction, we need to indirectly carry out the following SVD,
\begin{align}
	Q_{\nu_0\nu_1\nu_2\nu_3}\approx\sum_{k=1}^{D}U_{\nu_0\nu_1,k}s_{k}{V^{\dag}}_{k,\nu_2\nu_3},
\end{align}
with $D$ the bond dimension in the GATRG. In the block diagonal representation, this is expressed by
\begin{align}
	\begin{bmatrix}
		Q^{{\rm (even)}} & 0\\
		0 & Q^{{\rm (odd)}}
	\end{bmatrix}
	\approx
	\begin{bmatrix}
		U^{{\rm (even)}} & 0\\
		0 & U^{{\rm (odd)}}
	\end{bmatrix}
	\begin{bmatrix}
		s^{{\rm (even)}} & 0\\
		0 & s^{{\rm (odd)}}
	\end{bmatrix}
	\begin{bmatrix}
		V^{{\rm (even)}\dag} & 0\\
		0 & V^{{\rm (odd)}\dag}
	\end{bmatrix}.
\end{align}
Let $\tilde{k}$ be the fermion index for $k$. Then we assign $\tilde{k}=0(1)$ when $s_{k}$ comes from the matrix $s^{{\rm (even)}}(s^{{\rm (odd)}})$. In addition, the information of the fermion index $\tilde{k}$ can be encoded in the ordering of $k$:
\begin{align*}
\label{eq:f_index}
	k=\underbrace{1,\cdots,d}_{\tilde{k}=0},\underbrace{d+1,\cdots,D}_{\tilde{k}=1}.
\end{align*}
This means that $D$ largest singular values in $s$ consist of $d$ singular values in $s^{{\rm (even)}}$ and $D-d$ singular values in $s^{{\rm (odd)}}$. A similar ordering trick is also available in the initial tensor network representation.\footnote{Another ordering trick is demonstrated in Ref.~\cite{Akiyama:2020sfo}.} Each index in the initial tensor $T_{n;txyzt'x'y'z'}$ is composed of three integers, say $i=(i_1,i_2,i_3)$, running from $(0,0,0)$ to $(1,1,1)$. Note that $i_1$, $i_2$ and $i_3$ have been introduced via Eqs.~\eqref{eq:forward}, \eqref{eq:backward} and \eqref{eq:four-fermi}, respectively. Then the following mapping is practically useful:
\begin{align*}
	(0,0,0)\mapsto 1, ~~(1,1,0)\mapsto 2, ~~(0,0,1)\mapsto 3, ~~(1,1,1)\mapsto 4,\\
	(1,0,0)\mapsto 5, ~~(0,1,0)\mapsto 6, ~~(1,0,1)\mapsto 7, ~~(0,1,1)\mapsto 8. 
\end{align*}
As we have seen in Eq.~\eqref{eq:grassmann-order}, the third component, say $i_3$, does not affect the Grassmann parity in the initial tensor. As a result, the parity of the sum of the first two components, $i_1$ and $i_2$, corresponds to the Grassmann parity. Then the fermion index can be encoded in the ordering as
\begin{align*}
	\underbrace{1,~~2,~~3,~~4}_{{\rm (fermion~index)}=0},\underbrace{5,~~6,~~7,~~8}_{{\rm (fermion~index)}=1}.
\end{align*}
Thanks to this trick, Eq.~\eqref{eq:sign} is simplified with the corresponding fermion indices as
\begin{align}
	\sigma_{n+\hat{3},n}&=(-1)^{\tilde{z}(n)+\tilde{t}(n+\hat{3})\tilde{t}(n)+[\tilde{t}(n+\hat{3})+\tilde{x}(n+\hat{3})]\tilde{x}(n)+[\tilde{t}(n+\hat{3})+\tilde{x}(n+\hat{3})+\tilde{y}(n+\hat{3})]\tilde{y}(n)}\nonumber\\
	&~\quad(-1)^{\tilde{y}'(n+\hat{3})\tilde{y}'(n)+[\tilde{x}'(n+\hat{3})+\tilde{y}'(n+\hat{3})]\tilde{x}'(n)+[\tilde{t}'(n+\hat{3})+\tilde{x}'(n+\hat{3})+\tilde{y}'(n+\hat{3})]\tilde{t}'(n)}.
\end{align}

The last technique to be mentioned is the parallel computation, which reduces the execution time of the GATRG. The essence of this technique in the ATRG is demonstrated in Ref.~\cite{Akiyama:2020Dm}; the computational cost per process of tensor contraction is reduced from $O(D^9)$ to $O(D^8)$. As in Refs.~\cite{Akiyama:2020Dm,Akiyama:2020ntf}, we employ the randomized SVD (RSVD) in the swapping bond part. The accuracy of the RSVD is controlled by the oversampling parameter $p$ and $q$ iterations of QR decomposition. Under the block diagonal representation, we apply the RSVD with $p=4D$ and $q=D$ to each block matrix.

\section{Numerical results} 
\label{sec:results}
 
\subsection{Setup}
\label{subsec:setup}

We choose a large value of $g_0=32$ for the four-fermi coupling in Eq.~(\ref{eq:njl_lat}), because the FRG analysis in Ref.~\cite{Aoki:2017rjl} indicates the vanishing phase transition for smaller $g_0$. The partition function of Eq.~\eqref{eq:network} is evaluated, using the GATRG algorithm on a lattice up to the volume $V=L^4$ ($L=2^m, m \in \mathbb{N}$). We assume the periodic boundary conditions for $x$-, $y$-, $z$-directions and the anti-periodic boundary condition for $t$-direction.

Before investigating the restoration of the chiral symmetry at vanishing fermion mass, we check the efficiency of the GATRG algorithm by benchmarking with the NJL model in the heavy dense limit, which is defined as $m\to\infty$ and $\mu\to\infty$, keeping ${\rm e}^{\mu}/m$ fixed. The heavy dense limit gives us an opportunity to compare numerical results with the exact analytical ones.

\subsection{Heavy dense limit as a benchmark}
\label{subsec:heavydense}

In the heavy dense limit, the number density $\braket{n}$ and the fermion condensate $\braket{ {\bar \chi}(n)\chi(n)}$ at vanishing temperature can be derived analytically as
\begin{align}
\label{eq:hd_n}
	 \braket{n}=\Theta(\mu-\mu_c),
\end{align}
\begin{align}
\label{eq:hd_c}
	 \braket{ {\bar \chi}(n)\chi(n)}=\frac{1}{m}\Theta(\mu_c-\mu),
\end{align}
where $\Theta$ denotes the step function and $\mu_c=\ln(2m)$ \cite{Pawlowski:2013gag}. 

\begin{figure}[htbp]
	\centering
	\includegraphics[width=0.7\hsize,bb= 0 0 792 612]{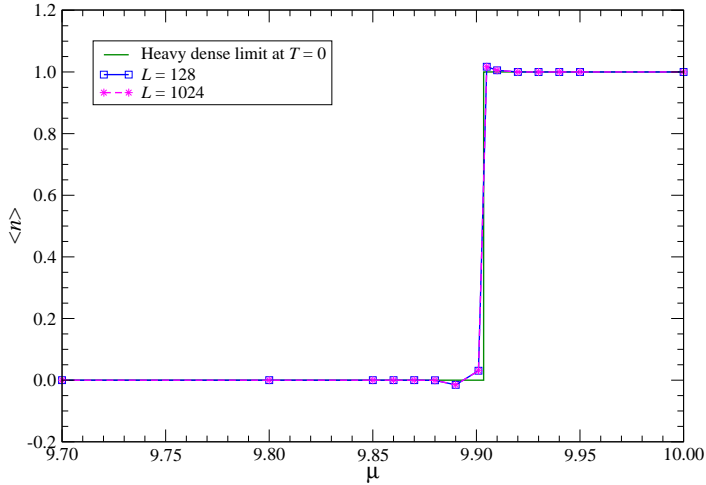}
  	\caption{Number density at $m=10^{4}$ and $g_0=32$ on $128^4$ and $1024^4$ lattices as a function of $\mu$ with $D=30$. $\Delta\mu=4.0\times10^{-3}$ in the vicinity of $\mu_c$. Green line denotes the step function in Eq.~\eqref{eq:hd_n}.}
  	\label{fig:hd_n}
\end{figure}

\begin{figure}[htbp]
	\centering
	\includegraphics[width=0.7\hsize,bb= 0 0 792 612]{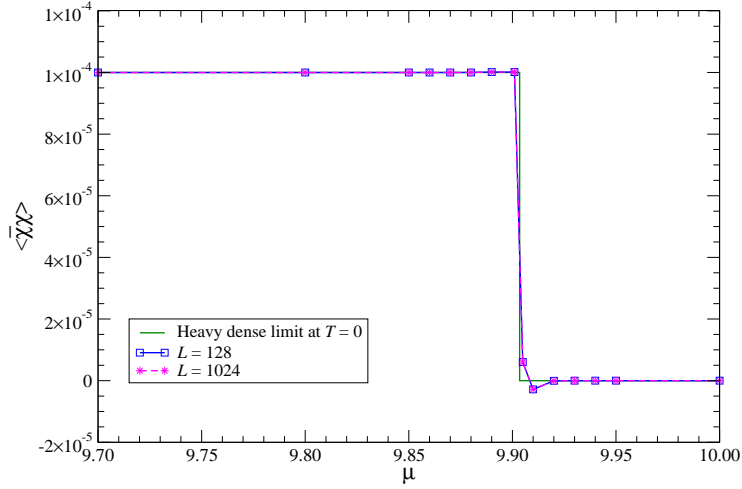}
  	\caption{Fermion condensate at $m=10^{4}$ and $g_0=32$ on $128^4$ and $1024^4$ lattices as a function of $\mu$ with $D=30$. Green line denotes the step function in Eq.~\eqref{eq:hd_c}.}
  	\label{fig:hd_c}
\end{figure}

Figures \ref{fig:hd_n} and \ref{fig:hd_c} show the numerical results for $\braket{n}$ and $\braket{ {\bar \chi}(n)\chi(n)}$ obtained by the GATRG algorithm choosing $m=10^4$ with $D=30$. 
The number density is calculated by the numerical derivative of the thermodynamic potential in terms of the chemical potential:
\begin{align}
\label{eq:nd_number}
	\braket{n}=\frac{1}{V}\frac{\partial \ln Z(\mu)}{\partial\mu}\approx\frac{1}{V}\frac{\ln Z(\mu+\Delta \mu)-\ln Z(\mu)}{\Delta \mu}.
\end{align}
In the vicinity of $\mu_c$, we have set $\Delta\mu=4.0\times10^{-3}$. The fermion condensate is also obtained via the numerical derivative of the thermodynamic potential in terms of $m$:
\begin{align}
	\left.\braket{ {\bar \chi}(n)\chi(n)}\right|_{m=10^4}=\left.\frac{1}{V}\frac{\ln Z(m+\Delta m)-\ln Z(m)}{\Delta m}\right|_{m=10^4}
\end{align}
with $\Delta m=1$. 
Since there is little difference between the $L=128$ and $1024$ results, the $L=1024$ lattice is sufficiently large to estimate the thermodynamic limit at vanishing temperature. The numerical results well reproduce the analytical ones, including the location of $\mu_c=\ln(2m)=9.903$, both for $\braket{n}$ and $\braket{ {\bar \chi}(n)\chi(n)}$ in the heavy dense limit. Note that the results quickly converge with respect to $D$; the difference between $\ln Z(D=25)$ and  $\ln Z(D=30)$ has been already suppressed less than $2.1\times10^{-3}\%$ in the vicinity of $\mu_c$. 

\subsection{Chiral phase transition}
\label{sec:aphase}

\begin{figure}[htbp]
	\centering
	\includegraphics[width=0.7\hsize,bb= 0 0 792 612]{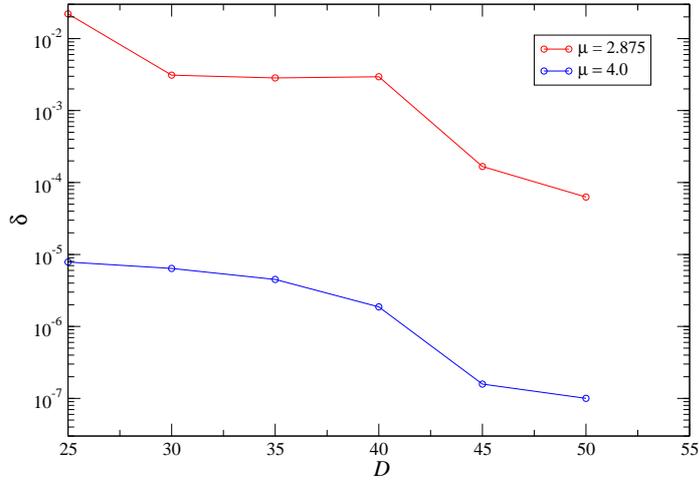}
  	\caption{Convergence behavior of thermodynamic potential as a function of $D$ on $V=1024^4$ with $m=0.01$.}
  	\label{fig:D_vs_F}
\end{figure}

Having confirmed the efficiency of the GATRG algorithm in the heavy dense limit, let us turn to the calculation with the light fermion masses. 
We first check the convergence behavior of the thermodynamic potential by defining the quantity
\begin{align}
	\delta=\left|\frac{\ln Z(D)-\ln Z(D=55)}{\ln Z(D=55)}\right|
\end{align}
on $V=1024^4$. In Fig.~\ref{fig:D_vs_F}, we plot the $D$ dependence of $\delta$  at $\mu=2.875$, which is near the phase transition point and $\mu=4.0$, which is in the dense region with the restored chiral symmetry, as we will see below. 
Although both of them are in the cold and dense region characterized with $\mu/T\sim O(10^3)$, where the Monte Carlo simulation should be severely hindered by the sign problem, good convergent behaviors are observed in the GATRG calculation; near the transition point, $\delta$ is reduced to about $10^{-4}$ up to $D=55$ and better convergence behavior, $\delta\lesssim10^{-7}$, is observed at $\mu=4.0$. Hereafter we present the results at $D=55$.

\begin{figure}[htbp]
	\centering
	\includegraphics[width=0.7\hsize,bb= 0 0 792 612]{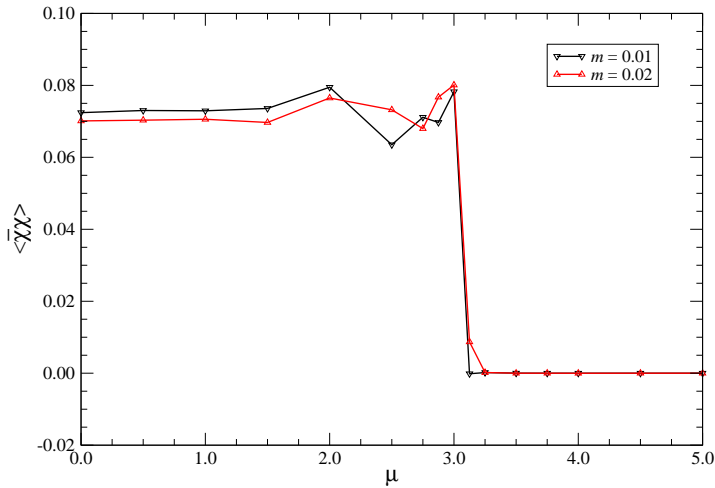}
  	\caption{Chiral condensate at $m=0.01$ and $0.02$ on $1024^4$ lattice as a function of $\mu$ with $D=55$.}
  	\label{fig:cc}
\end{figure}

\begin{figure}[htbp]
	\centering
	\includegraphics[width=0.7\hsize,bb= 0 0 792 612]{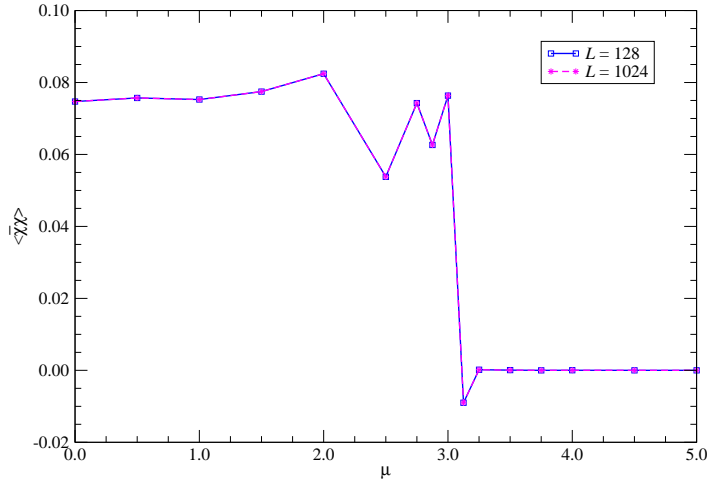}
  	\caption{Chiral condensate extrapolated in the chiral limit as a function of $\mu$ with $D=55$ on $128^4$ and $1024^4$ lattices.}
  	\label{fig:cc_m0}
\end{figure}

We investigate the chiral phase transition employing the chiral condensate $\braket{ {\bar \chi}(n)\chi(n)}$, as an order parameter, which is defined by
\begin{align}
\label{eq:def_cc}
	\braket{ {\bar \chi}(n)\chi(n)}=\lim_{m\to0}\lim_{V\to\infty}\frac{1}{V}\frac{\partial}{\partial m}\ln Z,
\end{align}
in the cold region. We calculate $\braket{ {\bar \chi}(n)\chi(n)}$ with the numerical derivative of thermodynamic potential and the chiral extrapolation with the corresponding results at finite mass in the thermodynamic limit.\footnote{It is possible to evaluate the chiral condensate with the impurity tensor method \cite{Yoshimura:2017jpk,MORITA201965}. Since Eq.~\eqref{eq:network} consists of eight types of tensor, there are eight configurations of an impurity tensor. Consequently, the computational cost is eight times larger than that of coarse-graining Eq.~\eqref{eq:network}. One can also evaluate the number density discussed below with the impurity tensor method, which requires four times larger cost than that to coarse-grain Eq.~\eqref{eq:network} .} In this study, the partial derivative in Eq.~\eqref{eq:def_cc} is numerically evaluated via
\begin{align}
	\frac{\partial}{\partial m}\ln Z\approx\frac{\ln Z(m+\Delta m)-\ln Z(m)}{\Delta m},
\end{align}
with $\Delta m=0.01$. In Fig.~\ref{fig:cc}, we plot the $\mu$ dependence of the chiral condensate at $m=0.01$ and 0.02 on the $L^4=1024^4$ lattice. The signals show slight fluctuations as a function of $\mu$ around the transition point.  Away from the transition point, we have found little response in $\braket{ {\bar \chi}(n)\chi(n)}$ to changes in mass. Figure~\ref{fig:cc_m0} presents the results in the chiral limit obtained by the chiral extrapolation with the data at $m=0.01$ and $0.02$ on two volumes of $L^4=128^4$ and $1024^4$. It is hard to find the difference between the $L=128$ and $1024$ results. This allows us to consider the $L=1024$ result to be essentially in the thermodynamic limit. We observe the discontinuity from a finite value to zero for the chiral condensate at $\mu_c= 3.0625\pm 0.0625$, which is a clear indication of the first-order phase transition. Note that enlarging the bond dimension $D$ is more essential than adding the data points at different fermion masses in order to increase the numerical accuracy around $\mu_c$ found in Fig.~\ref{fig:cc_m0}.

\subsection{Equation of state}
\label{subsec:eos}

\begin{figure}[htbp]
	\centering
	\includegraphics[width=0.7\hsize,bb = 0 0 792 612]{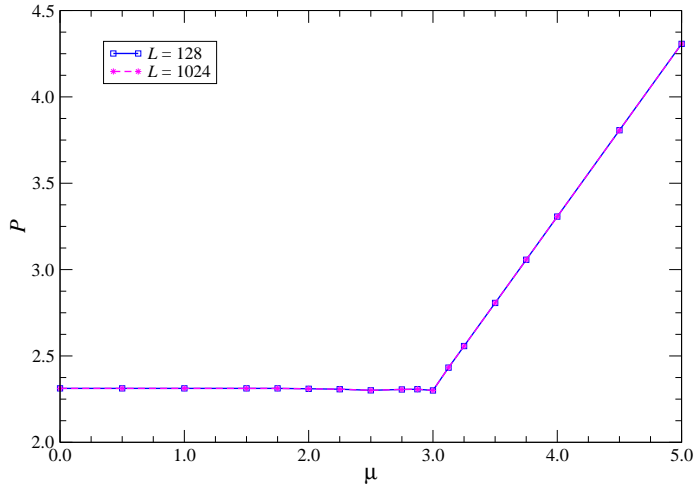}
   	\caption{Pressure at $m=0.01$ as a function of $\mu$ on $128^4$ and $1024^4$ lattices.}
  	\label{fig:pressure}
\end{figure}

The equation of state is a relation between the pressure and the particle number density. Here we present both results as functions of $\mu$, respectively.
In the thermodynamic limit, the pressure $P$ is directly obtained from the thermodynamic potential:
\begin{align}
	P=\frac{\ln Z}{V},
\end{align}
where the vast homogeneous system is assumed. In Fig.~\ref{fig:pressure}, we plot the $\mu$ dependence of the pressure at $m=0.01$. We find a kink behavior at $\mu_c= 3.0625\pm 0.0625$, where the chiral condensate shows the discontinuity. Note that the $m=0.02$ result shows little difference from the $m=0.01$ one.

\begin{figure}[htbp]
	\centering
	\includegraphics[width=0.7\hsize,bb = 0 0 792 612]{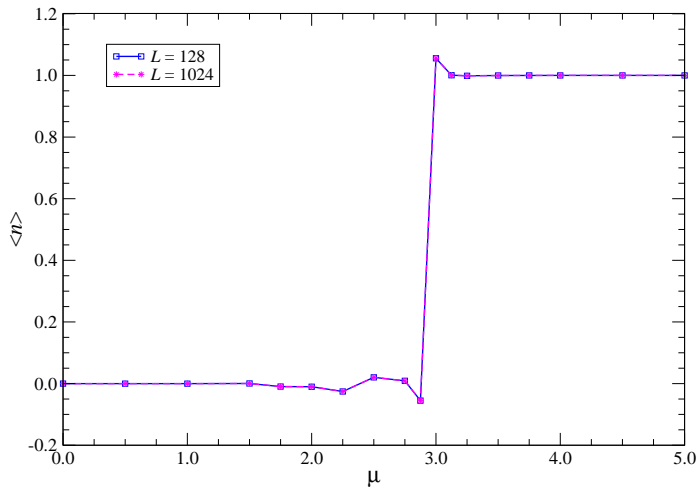}
  	\caption{Particle number density at $m=0.01$ as a function of $\mu$ on $128^4$ and $1024^4$ lattices.}
  	\label{fig:number_density}
\end{figure}

Fig.~\ref{fig:number_density} shows the $\mu$ dependence of the particle number density $\braket{n}$  obtained by Eq.~\eqref{eq:nd_number}. We observe an abrupt jump from $\braket{n}=0$ to $\braket{n}=1$ at $\mu_c= 2.9375\pm 0.0625$. This is another indication of the first-order phase transition. The small shift of $\mu_c$ compared to the cases of chiral condensate and pressure is attributed to the definition of the numerical derivative in Eq.~\eqref{eq:nd_number}.

\section{Summary and outlook} 
\label{sec:summary}

We have investigated the restoration of the chiral symmetry of the NJL model in the dense region at very low temperature, employing the Kogut--Susskind fermion action on the extremely large lattice of $V=1024^4$, which is in the thermodynamic limit at zero temperature, essentially. The first-order phase transition is clearly observed using the chiral condensate as an order parameter. At the critical chemical potential, we also find the jump of the number density. 

This is the third successful application of the TRG method to the $4d$ lattice theories, following the Ising model \cite{Akiyama:2019xzy} and the complex $\phi^4$ theory at finite density \cite{Akiyama:2020ntf}. This study is also the first application to the $4d$ fermionic system. As a next step, it would be interesting to determine the critical end point of this model.

\begin{acknowledgments}
Numerical calculation for the present work was carried out with the Oakforest-PACS (OFP) and the Cygnus computers under the Interdisciplinary Computational Science Program of Center for Computational Sciences, University of Tsukuba.
This work is supported in part by Grants-in-Aid for Scientific Research from the Ministry of Education, Culture, Sports, Science and Technology (MEXT)
(No. 20H00148).
\end{acknowledgments}

\bibliographystyle{JHEP}

\bibliography{formulation,algorithm,discrete,grassmann,continuous,gauge,for_this_paper}

\end{document}